Letter

# Optimizing QKD efficiency by addressing chromatic dispersion and time measurement uncertainty

Artur Czerwinski [a,*], Saeed Haddadi [b]

[a] Institute of Physics, Faculty of Physics, Astronomy and Informatics, Nicolaus Copernicus University in Torun, ul. Grudziadzka 5, 87-100 Torun, Poland
[b] Faculty of Physics, Semnan University, P.O. Box 35195-363, Semnan, Iran

ABSTRACT

In this paper, we present a Quantum Key Distribution (QKD) protocol that accounts for fundamental practical challenges, including chromatic dispersion, time measurement uncertainty, and dark counts. Our analysis provides a comprehensive framework for understanding the impact of these physical phenomena on QKD efficiency, offering practical strategies for enhancing the robustness and security of quantum communication systems in real-world applications. In particular, by manipulating the chirp parameter of single-photon wave packets, we demonstrate significant improvements in key generation rates and an extended range of secure communication.

## 1. Introduction

Quantum Key Distribution (QKD) is a cornerstone of quantum communication, offering a theoretically unbreakable method for securely transmitting cryptographic keys between distant parties [1]. Unlike classical encryption methods, QKD relies on the principles of quantum mechanics, where the security of the key is guaranteed by the fundamental laws of physics rather than computational complexity [2]. The key is typically encoded in the quantum states of single photons, making the precise manipulation and detection of these photons crucial for the successful implementation of QKD protocols [3,4].

However, practical implementations of QKD systems face several challenges that can compromise their efficiency and security. In real-world scenarios, the transmission of single photons through optical fibers or free-space channels is subject to various physical phenomena that distort the quantum states. Chromatic dispersion, for instance, causes temporal spreading of the photon wave packets, due to wavelength-dependent velocity in a dispersive medium [5–8]. As a result, the temporal width of the photon increases with the length of the dispersive quantum channel, requiring longer detection windows. Although this prevents signal loss, it also raises the probability of detecting noise, which can reduce the overall performance of the quantum communication protocols.

Other physical phenomena impact QKD protocols by introducing errors and causing losses [9,10]. Attenuation in the channel reduces the number of photons that reach the receiver, while detector imperfections, such as dark counts, introduce additional noise. Time measurement uncertainty further complicates the accurate timing and identification of photon arrival, impacting the overall key rate and security of the QKD system.

Recent advancements in QKD have significantly expanded its practical applications by addressing key challenges. It was demonstrated that secure communication over 833.8 km was achieved in fiber-based QKD using an optimized twin-field protocol, setting a new distance record and highlighting the potential for large-scale quantum-secure networks [11]. Additionally, improvements in QKD robustness were achieved with a high-speed system that compensates for channel polarization disturbances, maintaining a stable key rate over extended periods [12]. The importance of phase-control techniques and narrow-linewidth lasers for mitigating phase noise in twin-field QKD was also emphasized, particularly for real-world network deployments [13]. Further contributions have been made to the development of gated single-photon avalanche diodes (SPADs) for QKD, where the design of a specific low-pass filter (LPF) and a zero crossing discriminator significantly reduced time jitter at high gating frequencies, paving the way for practical QKD systems operating at higher clock rates [14].

Advancements in numerical methods have further strengthened QKD security and efficiency. Reliable techniques for calculating key rates under practical conditions have been developed, which are essential for assessing the performance of QKD protocols [15,16]. Moreover, a security framework addressing device imperfections was introduced, enhancing the viability of measurement-device-independent QKD for real-world applications [17]. At the same time, it was demonstrated that QKD protocols are vulnerable to practical security flaws, and device-independent QKD has been explored to address these issues, although it currently suffers from performance limitations [18]. Besides, it was shown that free-space quantum secure direct communication (QSDC), a related approach to QKD, has unique prospects for practical large-scale applications, especially in satellite-based systems [19]. These developments collectively illustrate the rapid progress in QKD and related technologies, bringing them closer to widespread adoption and securing future communication infrastructures.

* Corresponding author.
*E-mail addresses:* aczerwin@umk.pl (A. Czerwinski), haddadi@semnan.ac.ir (S. Haddadi).

Given practical challenges, it is still essential to explore strategies that can mitigate the effects of dispersion, attenuation, and other sources of error in QKD [20–23]. In particular, the manipulation of single photons, including their temporal characteristics, offers a promising avenue for optimizing QKD protocols. These approaches not only preserve the integrity of the quantum states but also improve the key generation rate, making QKD more robust against practical imperfections [24].

In this paper, we present a QKD protocol that incorporates the effects of chromatic dispersion, time uncertainty, and dark counts. Our analysis shows that the efficiency of the protocol can be improved by tailoring the chirp parameter of the photon wave packets. By carefully selecting this parameter, we can minimize the temporal broadening of the photons and optimize the timing accuracy of the detections. This optimization leads to higher key generation rates and enhances the overall performance of the QKD system, even in the presence of significant channel impairments.

The paper is organized as follows. Sec. 2 describes the theoretical foundations of the paper. It is divided into two parts: first, in Sec. 2.1, we review the fundamentals of chirped Gaussian pulses. Then, in Sec. 2.2, we introduce a QKD protocol that accounts for chromatic dispersion, time measurement uncertainty, and dark counts. The results of our numerical analysis are presented in Sec. 3. Finally, the paper concludes with a summary in Sec. 4.

## 2. Theoretical framework

### 2.1. Broadening of chirped Gaussian pulses

Let us consider a temporal mode of a single photon that can be described by a Gaussian temporal wave function (TWF)

$$\psi(t) = \frac{1}{\sqrt{\sqrt{2\pi}\sqrt{\sigma}}} \exp\left(-\frac{1+iC}{4\sigma^2}t^2\right), \tag{1}$$

where $C$ is a phase factor of the temporal mode, which is commonly referred to as a chirp parameter. The other parameter appearing in (1), $\sigma$, quantifies the initial temporal width of a photon.

For the wave function (1), we define a probability density function (PDF) following standard convention in quantum mechanics, i.e., $p(t) := \psi^*(t)\,\psi(t)$. One can verify that the PDF satisfies the following conditions

- $\int_{-\infty}^{\infty} p(t)\,dt = 1$,
- $\int_{-\infty}^{\infty} t\,p(t)\,dt = 0$,
- $\int_{-\infty}^{\infty} t^2\,p(t)\,dt = \sigma^2$.

These results signify that $p(t)$ is a normalized distribution with zero-mean, and $\sigma^2$ represents the variance of the PDF.

We describe the evolution of the temporal mode due to propagation through dispersive medium by using a propagator, $\mathcal{D}(t,\tilde{t},L)$ in the form [25–27]

$$\mathcal{D}(t,\tilde{t},L) = \frac{1}{\sqrt{4\pi i\beta L}} \exp\left(\frac{i(t-\tilde{t})^2}{4\beta L}\right), \tag{2}$$

where $L$ denotes the propagation distance and $\beta$ represents the group-velocity dispersion (GVD) parameter, in other words called the second-order dispersion parameter.

The propagator, $\mathcal{D}(t,\tilde{t},L)$, acts on the initial wave function (1) as follows

$$\psi_L(t) := \int_{-\infty}^{\infty} \mathcal{D}(t,\tilde{t},L)\,\psi(\tilde{t})\,d\tilde{t}. \tag{3}$$

By following (3), we arrive at the formula to describe the TWF after the propagation

$$\psi_{L(+)}(t) = \frac{1-i}{2^{3/4}\,\pi^{1/4}} \exp\left(\frac{1+iC}{4(C-i)L\beta-4\sigma^2}t^2\right) \times \sqrt{\frac{\sigma}{L\beta(1+iC)-i\sigma^2}} \quad \text{for} \quad \beta>0 \tag{4}$$

and

$$\psi_{L(-)}(t) = (-2\pi)^{-1/4} \exp\left(\frac{1+iC}{4(C-i)L\beta-4\sigma^2}t^2\right) \times \sqrt{\frac{\sigma}{i[(i-C)L\beta+\sigma^2)]}} \quad \text{for} \quad \beta<0. \tag{5}$$

For the modified TWF, we define a corresponding PDF as $p_{L(\pm)}(t) := \psi^*_{L(\pm)}(t)\psi_{L(\pm)}(t)$.

Recently, the framework of TWFs has been used to study the broadening of single photons caused by chromatic dispersion under various conditions [28]. Notably, it has been shown that negative chirp parameters can reduce the temporal broadening of a photon's shape. In this paper, we explore whether these negative chirp values can improve the performance of QKD.

### 2.2. QKD protocol involving chromatic dispersion and time uncertainty

We follow the BB84 QKD protocol [29], where the information bits 0 and 1 are encoded in the polarization degree of freedom. Alice, who wants to establish a secret key with Bob, selects randomly an orthonormal basis from two possibilities: $\{|H\rangle, |V\rangle\}$ or $\{|D\rangle, |A\rangle\}$. Then, the bit values 0 and 1 are encoded in the corresponding orthogonal states. A sequence of photons, which represents a binary code, is sent to Bob, who has to consider the temporal width of photons to properly detect the signal. Let us assume that the Bob's detector is characterized by a Gaussian profile [26,30]

$$\chi_d(t) = \frac{1}{\sqrt{2\pi\sigma_j^2}} \exp\left(-\frac{t^2}{2\sigma_j^2}\right), \tag{6}$$

where $\sigma_j$ stands for the measurement uncertainty (timing jitter). Then, the PDF of photon detection is given by a convolution

$$(p_L \star \chi_d)(t) := \int_{-\infty}^{\infty} p_L(\tau)\chi_d(t-\tau)d\tau \equiv \widetilde{p}_L(t), \tag{7}$$

where $p_L(t) \equiv p_{L(-)}(t)$ since we consider fiber-based communication with $\beta<0$. For example, standard single-mode fibers (SMF) typically exhibit anomalous dispersion at a wavelength of around 1550 nm, meaning that $\beta<0$ [31–33].

The expression (7) allows us to compute the probability of detecting a signal photon within a temporal window $\upsilon$, given by

$$p_{\text{sig}} = \int_{-\upsilon/2}^{\upsilon/2} \widetilde{p}_L(t)dt. \tag{8}$$

We assume that consecutive photons are separated by a temporal distance denoted by $\mathfrak{T}$. Then, Bob needs to consider the probability of detecting a photon that was sent in the previous time bin or in following bin from the viewpoint of selected detection window. By following (7), we obtain

$$\widetilde{p}_L^{\pm}(t) = \int_{-\infty}^{\infty} p_L(\tau \pm \mathfrak{T}) \chi_d(t-\tau) d\tau, \tag{9}$$

where '+/−' refers to photons following/preceding the signal. Overall, the probability of registering a wrong photon reads

$$p_w = \int_{-v/2}^{v/2} \widetilde{p}_L^{+}(t)dt \left(1 - \int_{-v/2}^{v/2} \widetilde{p}_L^{-}(t)dt\right) + \int_{-v/2}^{v/2} \widetilde{p}_L^{-}(t)dt \left(1 - \int_{-v/2}^{v/2} \widetilde{p}_L^{+}(t)dt\right), \tag{10}$$

where we intentionally exclude two-photon detection, because, in the event of getting two clicks in a single time slot, Bob can identify the issue and discard the affected data, ensuring it does not influence the key rate. We assume that Bob can identify this situation through the capabilities of his photon number resolving (PNR) detector. Recent advancements have shown that a PNR detection system can be achieved using a single superconducting nanowire single-photon detector (SNSPD) combined with threshold analysis of the electric waveform and ultra-high-resolution timing [34]. This setup enables Bob to distinguish between single-photon and multi-photon events effectively, allowing him to discard the data associated with double counts.

We then propose the following formula to calculate the probability of getting a single detection in a time window $v$ from photons sent by Alice

$$p_{\text{det}} = \eta(L) \left[p_{\text{sig}} + p_w(1 - \eta(L) p_{\text{sig}})\right], \tag{11}$$

where $\eta(L) = 10^{-\alpha L}$ denotes the channel transmittance for an attenuation coefficient $\alpha$.

Moreover, we need to notice that Bob's detection events stem from two sources: photons introduced into the channel by Alice and dark counts occurring in the Bob's detector. Contrary to the typical approach to dark counts [35], we do not disregard the probability of multiple detections. Instead, we use a Poisson distribution to accurately model the impact of dark counts. Let $\tilde{n}$ denote a random variable that represents the number of dark counts registered within the time window $v$. If the dark count rate (DCR) is denoted by $d$, the expected number of dark counts over the time window $v$ is given by $\langle \tilde{n} \rangle = dv$. Then, the random variable $\tilde{n}$ can be modeled by a Poisson distribution, i.e., $\tilde{n} \sim \text{Pois}(dv)$. Consequently, the probabilities of detecting zero and one dark count within the time window $v$ can be expressed as

$$p_{\text{zero}} = 1 - vd \quad \text{and} \quad p_{\text{one}} = dv(1 - dv), \tag{12}$$

where we use approximate formulas for the probabilities corresponding to zero and one event in the Poisson distribution.

Finally, we calculate the probability of generating a raw key bit as

$$p_{\text{raw}} = \frac{p_{\text{det}} p_{\text{zero}} + (1 - p_{\text{det}}) p_{\text{one}}}{2}. \tag{13}$$

The above equation represents the overall likelihood that Alice and Bob will obtain a key bit within the time window $v$. The fraction $1/2$ appearing in (13) is a result of the fact that Bob randomly chooses a basis for measurement and, for this reason, half of the detections will be discarded due to incompatible bases. Consequently, the quantum bit error rate (QBER) is given by

$$Q = \frac{1}{4} \cdot \frac{\eta(L) p_w (1 - \eta(L) p_{\text{sig}}) p_{\text{zero}} + (1 - p_{\text{det}}) p_{\text{one}}}{p_{\text{raw}}}, \tag{14}$$

where the fraction $1/4$ comes from the fact that only half of the "wrong" photons will cause errors in the secret key.

**Table 1**
Parameters used in the QKD analysis.

| DCR | $\sigma$ | $\beta$ | $\alpha$ | $\mathfrak{T}$ |
|---|---|---|---|---|
| 1 000 Hz | 10 ps | $-1.15 \times 10^{-26}$ s$^2$/m | 0.2 dB/km | 100 ps |

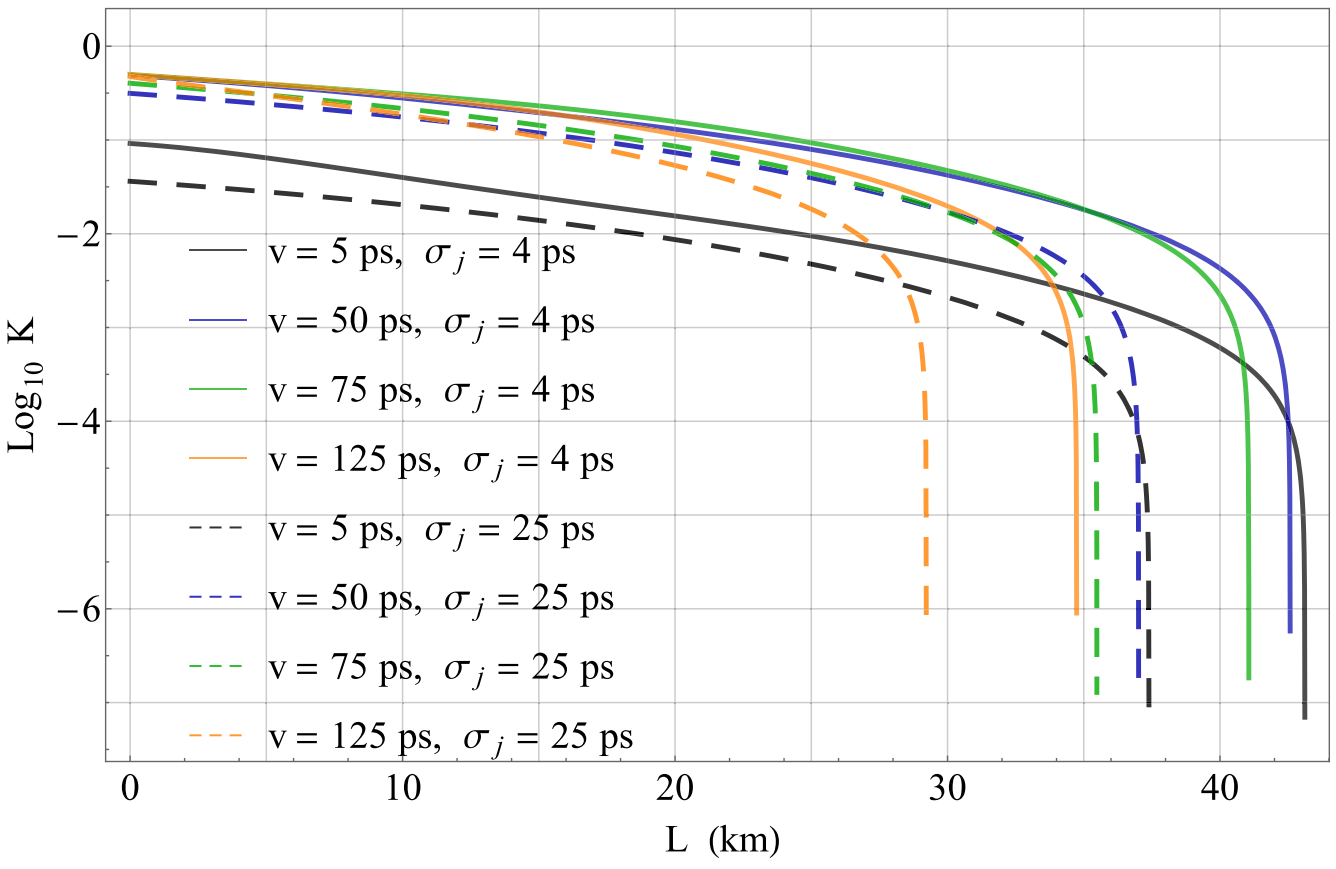

**Fig. 1.** Key generation rate for four values of the detection window and two jitters. Chirp parameter is fixed $C = 0$.

Finally, the key generation rate for BB84 can be computed as [18]

$$\mathcal{K} = \max\{0, \; p_{\text{raw}}(1 - 2\text{H}(Q))\}, \tag{15}$$

where $\text{H}(Q)$ denotes the binary entropy function for QBER, namely $\text{H}(Q) = -Q \log_2 Q - (1-Q)\log_2(1-Q)$.

The key generation rate is typically analyzed as a function of the communication distance $L$, and is often plotted against this variable. In this context, we also introduce the concept of the maximum distance of secure communication, denoted by $L_{\text{max}}$. This distance represents the point at which the key generation rate drops to zero, defined as

$$L_{\text{max}} \iff \mathcal{K}(L_{\text{max}}) = 0. \tag{16}$$

By incorporating the concept of the maximum distance, we can gain deeper insights into the efficiency and limitations of the QKD protocol under various conditions.

## 3. Key generation rate – results and analysis

From the key generation rate formula (15), one can expect an interplay between $\mathcal{K}$ and the other parameters such as chirp parameter $C$, detection window $v$, timing jitter $\sigma_j$ and so on. Notice that Table 1 presents the fixed values of selected parameters.

DCR is a key parameter that characterizes single-photon detection systems [36]. We assume that DCR= 1 kHz, a typical value for commercially available SNSPDs [37]. Although for state-of-the-art detectors DCR can be significantly reduced, our assumption implies that the framework is compatible with off-the-shelf technology.

Additionally, the transmission medium used in our model is a standard single-mode optical fiber, with attenuation and dispersion parameters, $\alpha$ and $\beta$, provided in Table 1, corresponding to $\lambda = 1550$ nm. These assumptions align the QKD model with current technological capabilities, ensuring its relevance to practical implementations.

Regarding time uncertainty, we do not fix the value of $\sigma_j$ but instead compare various values corresponding to different levels of detector technology. For off-the-shelf SNSPDs, this parameter is typically around 25 ps [38,39]. However, state-of-the-art detectors can achieve timing jitters as low as 4 ps at a wavelength of 1550 nm [40].

In Fig. 1, we illustrate the key rates versus the propagation distance $L$ for two values of detector jitter (time uncertainty) across four different detection windows. This figure presents the effect of time uncertainty and demonstrates the importance of optimizing the detection window. A wider window, such as $\upsilon = 125$ ps, increases the error probability, while a narrower window, such as $\upsilon = 5$ ps, reduces the chances of detecting a signal photon. Consequently, we identify $\upsilon = 50$ ps as the optimal detection window, irrespective of timing jitter. Therefore, we select $\upsilon = 50$ ps as the optimal value for generating the key rates shown in Figs. 2 to 4.

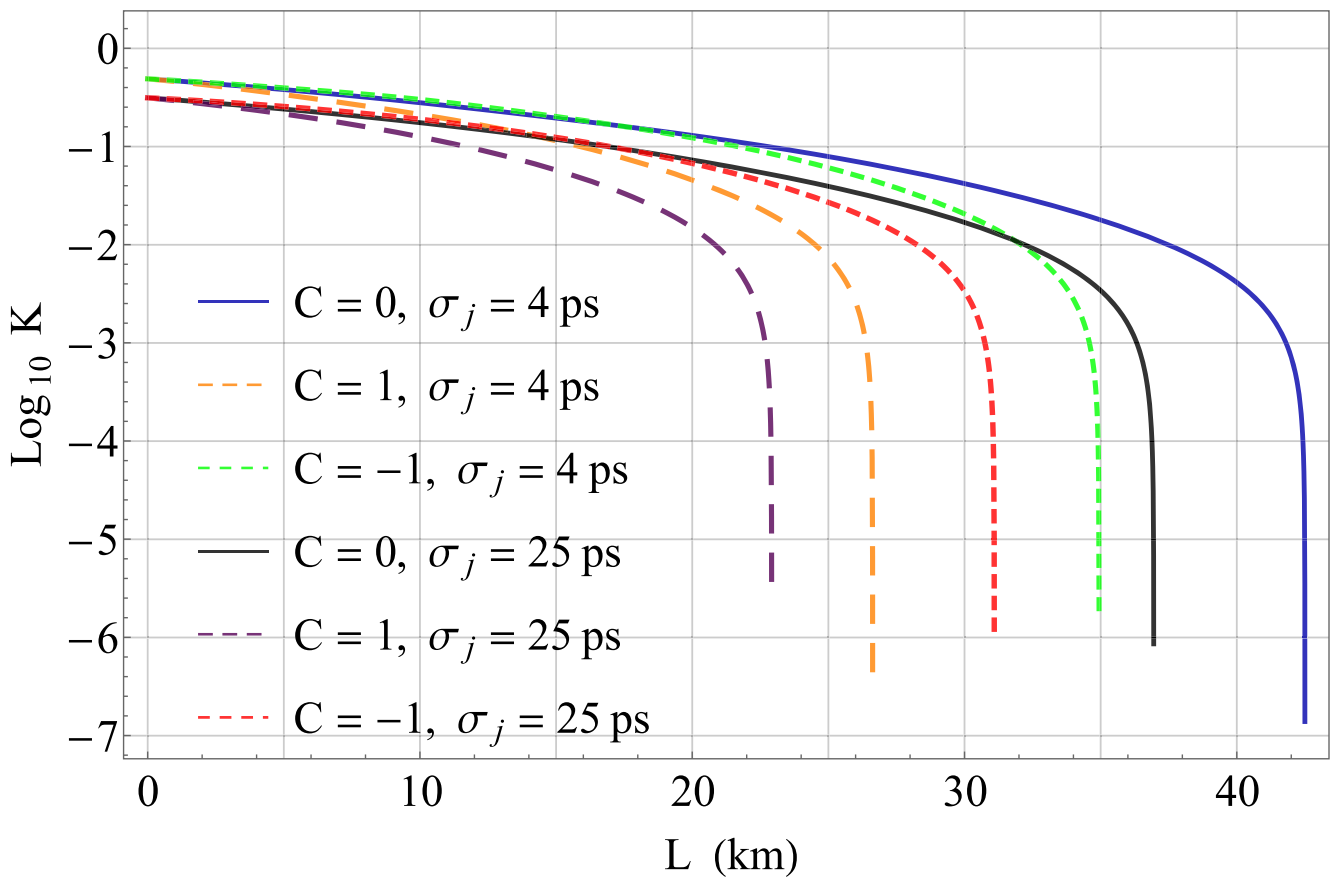

**Fig. 2.** Key generation rate for three values of the chirp parameter ($C = -1, 0, 1$) and two jitters. Detection window is fixed $\upsilon = 50$ ps.

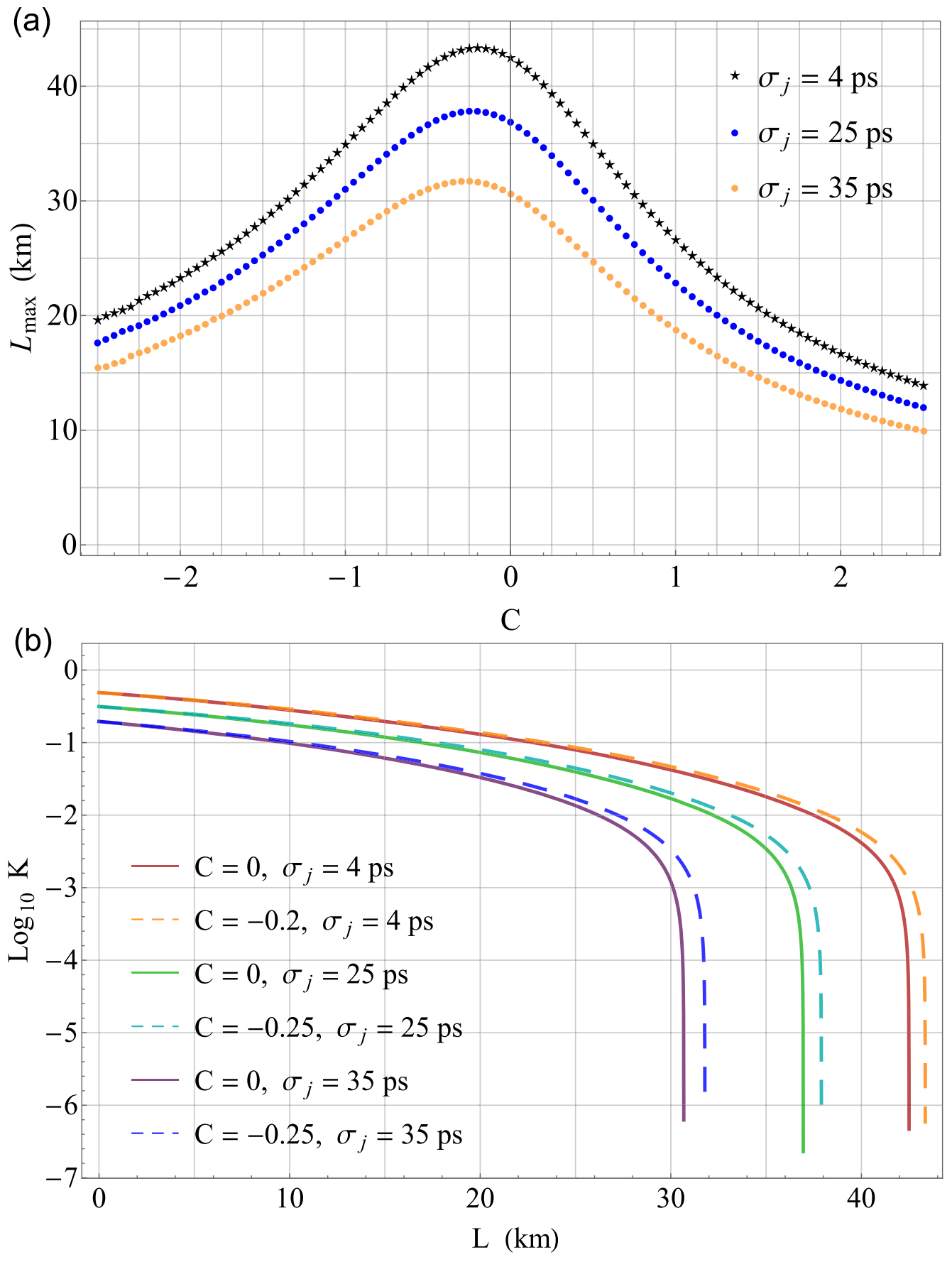

**Fig. 3.** (a) Maximum distance of secure quantum communication for three values of the detector's jitters, (b) Key generation rates optimized by negative chirp parameters. In both plots, detection window is fixed $\upsilon = 50$ ps.

Fig. 2 presents a comparison of key rates for two values of detector jitter ($\sigma_j = 4$ ps and $\sigma_j = 25$ ps) and three chirp parameters, $C = -1$, $C = 0$, and $C = 1$. The figure reveals that both $C = 1$ and $C = -1$ lead to a reduction in key rates compared to $C = 0$. However, the decrease in key rate is less pronounced for $C = -1$. Additionally, it is important to note that higher time uncertainty can lead to better key rates compared to lower jitter, depending on the chirp parameter values. This phenomenon is evident when comparing the dashed red line (corresponding to $C = -1$ and $\sigma_j = 25$ ps) with the dashed orange line (representing $C = 1$ and $\sigma_j = 4$ ps). This observation highlights the significant impact of the chirp parameter on key rates and the maximum range of QKD, indicating its potential to mitigate the effects of time uncertainty.

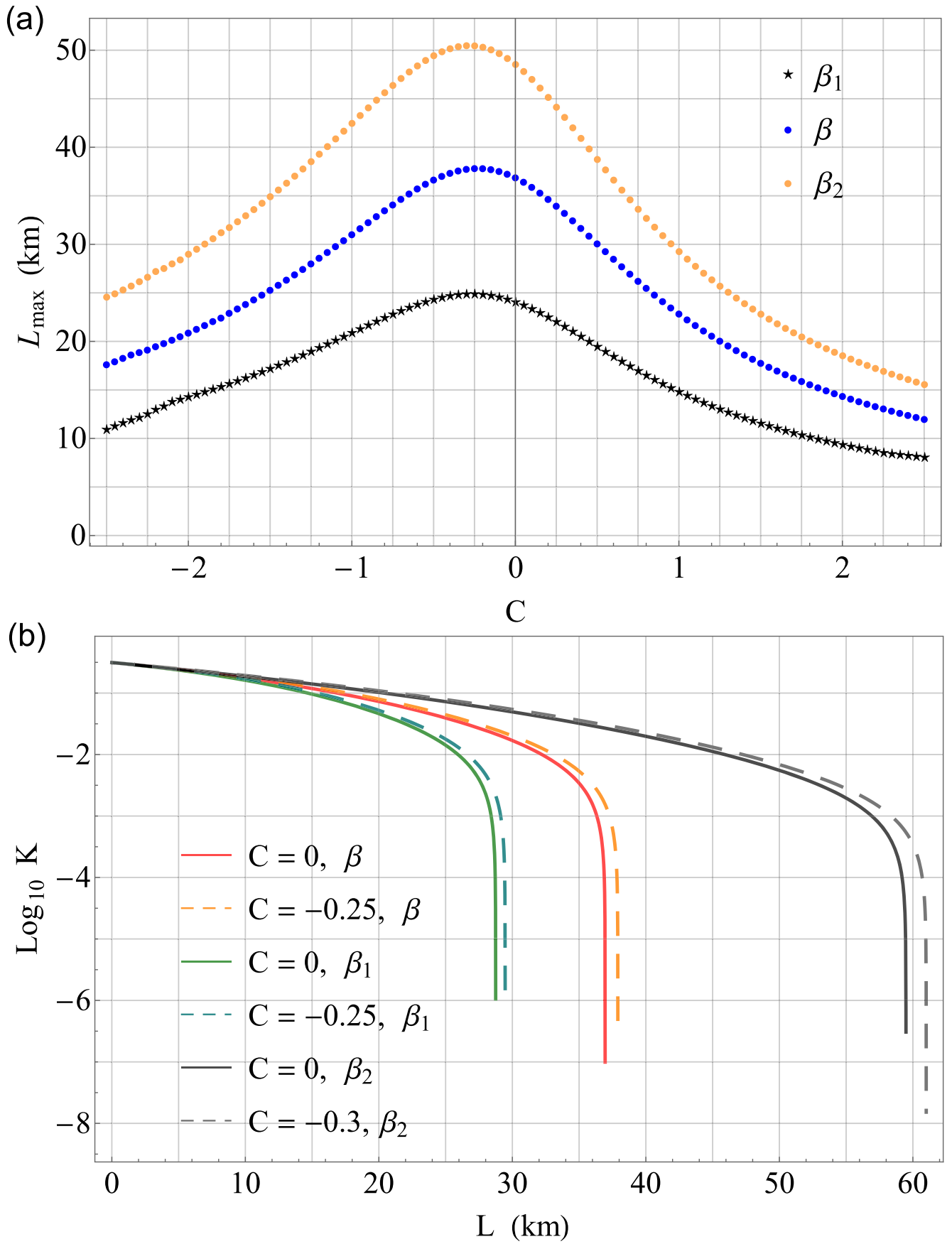

**Fig. 4.** (a) Maximum distance of secure quantum communication for three values of the GVD parameter, (b) Key generation rates optimized by negative chirp values for different GVD parameters. Detection window and jitter are fixed with $\upsilon = 50$ ps and $\sigma_j = 25$ ps.

The optimization of QKD efficiency is depicted in Figs. 3 and 4. The difference between the two figures is that Fig. 3(b) shows results for three different jitter values, while Fig. 4(b) compares a fixed $\beta$ with two other GVD parameters, $\beta_1$ and $\beta_2$. Apart from $\beta$ (as given in Table 1), we use $\beta_1 = -1.5 \times 10^{-26}$ s$^2$/m and $\beta_2 = -0.7 \times 10^{-26}$ s$^2$/m for comparison.

First, we present the maximum distance of secure communication, denoted by $L_{max}$ (16), as a function of the chirp parameter $C$ in Figs. 3(a) and 4(a). These figures clearly demonstrate the presence of a peak, indicating that selecting a chirp parameter $C$ slightly negative and close to zero improves QKD efficiency. This observation allows us to precisely identify the optimal values of $C$ that maximize the secure communication distance. The values of $C$ corresponding to the peak in the maximum secure communication distance are expected to optimize the overall QKD protocol.

To further explore optimization of QKD, we plot the key generation rate corresponding to these optimal values of $C$, as displayed in Figs. 3(b) and 4(b). The optimal values of the chirp parameter are found to lie between $C = -0.3$ and $C = -0.2$, and the corresponding key generation rates are compared to those with $C = 0$ (unchirped pulses) in Figs. 3(b) and 4(b). In all scenarios illustrated in Figs. 3 and 4, a signifi-cant improvement in QKD efficiency is observed when using a negative chirp parameter, compared to an unchirped pulse.

Additionally, in Fig. 4, we see that the parameter $\beta_2$, which features lower dispersion, leads to a higher key rate, as anticipated. However, the improvement when adjusting the chirp parameter from $C = 0$ to $C = -0.3$ is more pronounced for $\beta_2$ compared to $\beta$ and $\beta_1$. This indicates that the negative chirp parameter has a greater impact on enhancing the key rate in scenarios with lower dispersion. Therefore, by carefully controlling both the chirp parameter and the dispersion values, we can effectively mitigate the negative effects of chromatic dispersion on QKD and significantly improve the key generation rate.

## 4. Discussion and conclusions

The findings of this study present a significant contribution to the understanding and optimization of QKD protocols, particularly in the context of practical challenges such as chromatic dispersion, time measurement uncertainty, and dark counts. By investigating the role of the chirp parameter in the temporal mode of single photons, we have demonstrated that careful tuning of this parameter can substantially enhance the performance of QKD protocols. This optimization not only improves key generation rates but also extends the maximum secure communication range, even in the presence of substantial channel impairments.

One of the crucial insights from our work is the discovery that a slightly negative chirp parameter, close to zero, provides the best results in terms of key generation efficiency. Specifically, by plotting the maximum distance of secure communication, we were able to accurately determine the optimal chirp values, which outperform both unchirped pulses and other chirp configurations. This result suggests that QKD systems can be made more resilient to the detrimental effects of chromatic dispersion and timing jitter by fine-tuning the chirp parameter of the transmitted photons. This approach can be particularly effective in fiber-based communication systems where dispersion is a significant limiting factor.

Moreover, our analysis shows that the interplay between different physical parameters, such as GVD and timing jitter, can be effectively managed by adjusting the chirp parameter. For instance, the key generation rate improvements observed when using the optimal values of the chirp parameter were more pronounced in scenarios involving lower dispersion, which is typical for advanced optical fibers. This indicates that the chirp parameter can be used to mitigate the effects of both dispersion and timing uncertainties, leading to more robust and efficient QKD implementations.

In addition to enhancing key rates, our results also highlight the potential of the chirp parameter as a tool for mitigating the impact of environmental and system-induced uncertainties in quantum communication channels. The ability to manipulate the temporal characteristics of single photons in this manner opens up new avenues for optimizing QKD protocols under real-world conditions, where perfect control over every aspect of the communication channel is often unattainable.

These findings have important implications for the design and deployment of future QKD systems. By providing a practical method to improve QKD performance through chirp optimization, this study contributes to the ongoing efforts to transition quantum cryptography from theoretical research to widespread practical application. The strategies outlined here could be integrated into existing QKD systems with minimal modifications, thereby enhancing their security and efficiency without requiring significant changes to current infrastructure.

Our analysis could be extended to QKD protocols involving weak coherent pulses, which have garnered significant attention in recent years for their ability to achieve communication ranges in the hundreds of kilometers. This extension would further enhance the applicability of our findings to a broader range of QKD technologies.

In conclusion, this work not only advances our theoretical understanding of the factors influencing QKD performance but also offers concrete, practical guidelines for improving the robustness and efficiency of quantum communication systems. As quantum cryptography continues to evolve, the insights gained from this study will be crucial in overcoming the practical challenges that currently limit the widespread deployment of QKD technologies. Future research could explore the application of chirp optimization in different types of quantum channels, including free-space and satellite-based QKD, further broadening the scope and impact of these findings.

## CRediT authorship contribution statement

**Artur Czerwinski:** Writing – review & editing, Writing – original draft, Visualization, Validation, Supervision, Software, Resources, Project administration, Methodology, Investigation, Formal analysis, Conceptualization. **Saeed Haddadi:** Writing – review & editing, Writing – original draft, Formal analysis.

## Declaration of competing interest

The authors declare that they have no known competing financial interests or personal relationships that could have appeared to influence the work reported in this paper.

## Data availability

No data was used for the research described in the article.